\documentclass[fleqn,usenatbib]{mnras}
\usepackage{multirow}
\usepackage{rotating}
\usepackage{colortbl}
\usepackage{color}
\usepackage{amsmath}
\usepackage{amsfonts}
\usepackage{verbatim}
\usepackage{scalefnt}
\usepackage[percent]{overpic}
\usepackage[T1]{fontenc} 
\usepackage{aecompl} 
\usepackage{ulem}
\usepackage{cleveref}
\usepackage{numprint}

\usepackage{float}

\def\lsim{\mathrel{\rlap{\lower 3pt \hbox{$\sim$}} \raise 2.0pt \hbox{$<$}}}
\def\gsim{\mathrel{\rlap{\lower 3pt \hbox{$\sim$}} \raise 2.0pt \hbox{$>$}}}

\newcommand{\comments}[1]{} 

\title[Priorities in gravitational waveforms]{Priorities in gravitational waveforms for future space-borne detectors: vacuum accuracy or environment?}

\author[L. Zwick et al.]{Lorenz Zwick,\thanks{E-mail: zwicklo@ics.uzh.ch} Pedro R. Capelo and Lucio Mayer\\
Center for Theoretical Astrophysics and Cosmology, Institute for Computational Science, University of Zurich,\\
Winterthurerstrasse 190, CH-8057 Z{\"u}rich, Switzerland}

\date{Accepted XXX. Received YYY; in original form ZZZ}

\pubyear{2023}

\begin{document}

\label{firstpage}

\pagerange{\pageref{firstpage}--\pageref{lastpage}}

\maketitle


\begin{abstract}
In preparation for future space-borne gravitational-wave (GW) detectors, should the modelling effort focus on high-precision vacuum templates or on the astrophysical environment of the sources? We perform a systematic comparison of the phase contributions caused by 1) known environmental effects in both gaseous and stellar matter backgrounds, or 2) high-order post-Newtonian {(PN)} terms in the evolution of mHz GW sources {during the inspiral stage of massive binaries}. We use the accuracy of currently available analytical waveform models as a benchmark {value, finding} the following trends: the largest unmodelled phase contributions are likely environmental rather than PN for binaries lighter than $\sim 10^7/(1+z)^2$~M$_{\sun}$, where $z$ is the redshift. Binaries heavier than $\sim 10^8/(1+z)$~M$_{\sun}$ do not require more accurate {inspiral} waveforms due to low signal-to-noise ratios (SNRs). For high-SNR sources, environmental {phase contributions} are relevant at low redshift, while high-order vacuum templates are required at $z \gsim 4$. Led by these findings, we argue that including environmental effects in waveform models should be prioritised in order to maximize the science yield of future mHz detectors.
\end{abstract}



\vspace{-10pt}
\section{Challenges and opportunities in the millihertz band}\label{sec:introduction}

From the early post-Newtonian (PN) results \citep[see, e.g.][]{1985damour,1991damour,1998jaranowski,1999jaranowski}, formalisms such as the effective-one-body \citep[see, e.g.][]{1999buonanno,2001damoureob,2010barack,2012akcay,2014pan,2020ossokine} {and} self-force perturbation theory \citep[see, e.g.][]{1973Teukolsky,1997Quinn,2004burko,2011Gralla,2012lackeos}{, combined with advances in numerical relativity \citep[see, e.g.][]{2005pretor,ajith,purrer,pfeffer}\hypersetup{citecolor=blue},} have pushed the validity of analytical waveform templates to higher and higher orders. In combination with Bayesian inference techniques, they have made it possible to extract {an impressive amount of information} from gravitational-wave (GW) events detected by the Laser Interferometer GW Observatory \citep[LIGO; see, e.g.][]{abbot,2019chatzi,2020shaw,2021islam}. Now, the promise of space-borne mHz detectors in the early 2030s is breathing new life into the waveform modelling effort. Sources of GWs in the mHz band present ulterior challenges with respect to the ones routinely detected by ground-based observatories. Firstly, {crucial} parameters such as mass and mass ratio can vary by several orders of magnitude. Secondly, signal-to-noise ratios (SNRs) of several thousands are expected for appropriate sources \citep[see, e.g.][]{2017lisa}. These effects combined greatly increase the accuracy and breadth {demands of the required} waveform templates, motivating several research programmes with the aim of { developing more sophisticated analytical approximations as well as expanding the parameter space in which numerical relativity is viable \citep[see, e.g.][for some recent work]{2022lousto,2022Nagar}.}\hypersetup{citecolor=blue} {As an interesting recent example, template formalisms based on the PN and post-Minkowskian series have benefited from a recent influx of particle physics techniques \citep[see, e.g.][]{2019bern,2021mogull,2022buonanno}\hypersetup{citecolor=blue}, which are rapidly succeeding in computing high-order terms.}

A further, crucial difference between Hz and mHz binary sources is that the latter are more likely to be affected by their astrophysical environment. The presence of gas and massive third bodies can influence the source's evolution within the mHz band, confounding expectations based on vacuum templates, limiting the effectiveness of parameter estimation {(and thus tests of general relativity)}, and introducing spurious biases (see, e.g. \citealt{2008barausse}\hypersetup{citecolor=blue}; \citealt{2013gair,2014barausse,2020chen}; \citealt{2020caputo}\hypersetup{citecolor=blue}). Conversely, detecting these deviations represents a unique opportunity to measure properties of the source's environment (from constraining accretion disc physics to detecting dark matter and exoplanets; see, e.g. \citealt{1993chakrabarti,2017inayoshi}\hypersetup{citecolor=blue}; \citealt{2019tamanini,2019andrea,2020cardoso,2021andrea,2022zwick,2022speri,2022coogan}). A question is often posed in the context of the scientific groundwork required for missions such as the  Laser Interferometer Space Antenna (LISA; \citealt{roadmap2019}\hypersetup{citecolor=blue}; \citealt{2019lisa,2022lisaastro}) and TianQin \citep[][]{2016tianqin,2021tian}: {\it what is the relative importance of the environment with respect to the vacuum evolution of a GW source?} In this work, we {aim to to provide a simple but general framework to estimate the relative importance of environmental effects without needing to specify any particular waveform model.} {We apply this framework to analyse massive black hole (BH) binaries of varying mass and mass ratio, since they are often considered to be largely unaffected by their environments while radiating GWs in the mHz band.}

\vspace{-10pt}
\section{Comparing the phase of gravitational waves}\label{sec:effectivePN}\subsection{Vacuum waveforms}

The ability of a waveform template to accurately match the phase of a real signal is a crucial benchmark with regards to its accuracy. Over the course of an observation, the total phase $\phi$ of a {GW is} determined by the source's frequency evolution, $\dot{f}$. It reads \citep[see, e.g.][]{1994cutler}

\begin{align}
    \label{eq:totalphase}
    \phi = 2 \pi\int_{f_{\rm{min}}}^{f_{\rm{max}}}\frac{f_{\rm r}'}{\dot{f}\left(f_{\rm r}'\right)}{\rm d}f',
\end{align}

\noindent where $f$ is the observed GW frequency, $f_{\rm{r}} = f(1+z)$ is the rest frame frequency at a redshift $z$, $f_{\rm{min}}$ is the frequency at which the GW source enters a detector band, and $f_{\rm{max}}$ is the maximum frequency reached within the observation window. Sophisticated waveform models expand upon the historical result derived in \cite{1963peters} by including contributions from higher-order GW modes as well as PN corrections in the frequency evolution \citep[see, e.g.][]{2014blanchet}:

\begin{align}
    \label{eq:fdot}
    \dot{f}={\frac{q}{(1+q)^2}}\frac{96 \,\pi^{8/3}}{5}\frac{f_{\rm r}^{11/3} G^{5/3} M^{5/3}}{c^5}\left(1 + \mathcal{O}{\left[\left( \frac{v}{c} \right)^2\right]} \right),
\end{align}

\noindent where $M$ is the binary's total mass, $q$ its mass ratio, $v$ its characteristic orbital velocity, $c$ the speed of light in vacuum, and $G$ Newton's constant, and we assume quasi-circular orbits. Corrections to the leading order expression proportional to powers of $(v/c)^{2n}$ correspond to the $n$-th \textit{relative} PN order. {For the purposes of our phenomenological analysis of environmental effects, we make the two following assumptions regarding vacuum waveforms:}

\begin{itemize}

    \item {The phase accuracy of a waveform template is a proxy for its capacity to recover source parameters without bias. We thus neglect the merger and ring-down contributions to the SNR of a GW signal.\footnote{{Presumably, the additional SNR of the merger and ring-down signal might make it easier to distinguish environmental effects.}}}
    
    \item {The phase accuracy of any waveform template can be translated into an equivalent accuracy in terms of PN orders, regardless of the original formalism used to construct the template.}
    
\end{itemize}

{Both of these assumptions are motivated throughout the inspiral phase of the binary source, where environmental effects are most likely to be significant \citep[see, e.g.][]{2014blanchet}. Therefore, we deem a waveform template to be} accurate to the $n$-th PN order if it can match the phase of a true signal up to an error smaller than the next PN contribution $\sim (v/c)^{2n + 1}$, {as defined, e.g. by Eqs~(\ref{eq:fdot}--\ref{eq:fdotphenom})}. {In this language}, current state-of-the-art waveform models range in their {phasing} accuracy depending on several simplifying assumptions such as spin alignment, mass ratio{,} or lack of eccentricity, going as high as $\sim$20-PN \citep[see, e.g.][]{2015fujita,2020Munna}. In the case of a generic LISA source with arbitrary spins, moderate mass ratios ($1\lsim q \lsim 0.01$), and small eccentricities, purely analytical methods have achieved 4-PN accuracy, {while 5-PN accuracy has} been achieved by calibrating against numerical relativity simulations. {A general overview of recent literature leads us to set a reasonable benchmark of 5-PN phasing precision to be the current standard for waveform templates (based on several works;  see, e.g. \citealt{2019messina}; \citealt{2021Huber}\hypersetup{citecolor=blue}; \citealt{2021khalil,2021cho,2022Nagar,2022cho,2022blum}; \citealt{2022Chattaraj}\hypersetup{citecolor=blue}). While such a benchmark is arbitrary (and is bound to change in the following years), it will serve as a useful comparison tool to assess the current state of the field.}

In order to {mimic vacuum waveform templates of arbitrary precision}, we model the phase evolution of a source using a phenomenological form for Eq.~\eqref{eq:fdot}, based on the PN series as well as dimensional arguments \citep[see also][]{2021ii2}. It reads

\begin{align}
    \label{eq:fdotphenom}
    \dot{f}=\frac{{q\,}96 \,\pi^{8/3}}{5{(1+q)^2}}\frac{f^{11/3} G^{5/3} M^{5/3}}{c^5}\left(1 + \sum^{j_{\rm{max}}}_{j=0} A_{ j}\left(\frac{v}{c} \right)^j \right),
\end{align}

\noindent where {the exponent $j$ denotes the $(n/2)$-th effective PN order and we arbitrarily set all dimensionless coefficients $A_j = 1$, thus only preserving the physical $(v/c)$ scaling information \citep[see, e.g.][]{2020cardosomaselli}\hypersetup{citecolor=blue}.} {Note that, while Eq.~\eqref{eq:fdotphenom} is only a crude simplification, there are several fundamental uncertainties in the modelling of environmental effects which will overshadow any loss of precision due to setting $A_j = 1$. Note also that we are assuming that the phase evolution can be well described by orbit-averaged equations, losing information regarding the initial true anomaly of the source, an important parameter in many waveform models.}

Even in the case of vacuum sources, the accuracy required to extract a maximal amount of information is bounded by the inherent SNR limitations of GW detectors. {A simple approximation of} the SNR of a GW event can be {found by using} the following formula \citep[see, e.g.][]{2016Klein}:

\begin{align}
    \label{eq:SNR}
    \rm{SNR}=\sqrt{2 \cdot 4\int^{{f}_{\text{max}}}_{{f}_{\text{min}}}{\rm d}{f}^\prime\frac{h_{\rm{c}}^2({f}^\prime)}{S_{\rm{t}}({f}^\prime){f}^{\prime2}}},
\end{align}

\noindent where $S_{\rm t}$ is the detector's power spectral density and $h_{\rm c}$ the source's characteristic strain. For the former, we take the LISA specifications as reported in \cite{2019robson}. The latter reads

\begin{align}
    \label{eq:charstrain}
    h_{\rm c} = \frac{q}{(1+q)^2}\sqrt{N_{\rm c}} \frac{8 \pi ^{2/3}}{\sqrt{10}} \frac{G^{5/3}M^{5/3} f_{\rm r}^{2/3}}{D_{\rm{l}}(z)c^4},
\end{align}

\noindent where $D_{\rm l}$ is the source's luminosity distance and $N_{\rm c}$ is the number of cycles it completes at a given rest frame frequency. For an observation window of 4~yr, the latter reads

\begin{align}
    \label{eq:cycles}
    N_{\rm c} = \rm{min}\left(f_{\rm{r}} \times 4\, \rm{yr}, \frac{f_{\rm r}^2}{\dot{f}}\right).
\end{align}

{In order to define whether a small phase shift $\delta \phi$ is distinguishable from noise, we consider the following criterion, commonly used in more qualitative treatments of environmental effects on GWs \citep[see, e.g.][]{kocsis}:}

\begin{align}
    \label{eq:deccriterion}
     \sqrt{2 \cdot 4\int^{{f}_{\text{max}}}_{{f}_{\text{min}}}{\rm d}{f}^\prime\frac{h_{\rm{c}}^2({f}^\prime)}{S_{\rm{t}}({f}^\prime){f}^{\prime2}}\left( 1 - \cos\left(\delta \phi \right) \right)} > \delta \rm{SNR},
\end{align}

\noindent where $\delta$SNR is a detectability threshold, customarily chosen to be equal to 8. With the aid of Eq.~\eqref{eq:SNR} and {assuming} a constant phase shift, Eq.~\eqref{eq:deccriterion} can be rearranged into a useful form,

\begin{align}
    \label{eq:mindeph}
    \delta \phi > \arccos \left(1-\left(\frac{\delta \rm{SNR}}{\rm{SNR}}\right)^2\right),
\end{align}

\noindent which essentially states that the phase of a GW signal can be reconstructed with an accuracy of $\sim 2 \pi/$SNR \citep[see also, e.g.][]{2021katz}. {Phenomenologically,} waveform templates are therefore only required to achieve an accuracy comparable to the limit imposed by Eq.~\eqref{eq:mindeph}, since any smaller contribution to the phase of the real signal would be washed away by noise. Adopting a more accurate criterion would likely decrease the sensitivity to account for {the phase's accumulation rate along with degeneracies and other subtleties of more accurate GW data analysis.} {Note also, that Eqs~(\ref{eq:SNR}--\ref{eq:cycles}) are technically only valid at Newtonian order \citep[][]{2019Mangiagli}\hypersetup{citecolor=blue}. However, they suffice for the purposes of this work as they only serve to produce a reference SNR value, which we apply as a detectability criterion equally for both PN and environmental phase shifts: a crude but fair comparison in line with the phenomenological form of Eq.~\eqref{eq:fdotphenom} and the intrinsic uncertainties of astrophysical environmental effects.}

\vspace{-10pt}
\subsection{A sample of environmental effects}\label{sec:env}

Environmental influences can introduce additional terms that modify the frequency evolution of a source of GWs. In this work, we consider a \textit{minimal model} for three types of environmental effects that are considered typical in the astrophysical setting of LISA sources, i.e. we only consider a simple phase contribution. Richer environmental signatures can also be produced (see, e.g. \citealt{2022zwick} {and} \citealt{2022cardoso}\hypersetup{citecolor=blue} for gas{-}embedded binaries, or \citealt{2019alejandro,2021alejandro} for sources with a peculiar velocity), but are beyond the scope of this work. {The influence of additional environments} such as dark matter or other baryonic fields has also been explored in the literature (see, e.g. \citealt{Macedo2013}\hypersetup{citecolor=blue}; \citealt{2013kazunari}; \citealt{2016cardoso}\hypersetup{citecolor=blue}; \citealt{2022cole,2022baumann}).\newline

\textbf{Gas torques} can act on a GW source by transferring energy and angular momentum between the binary and the surrounding gas. The presence of gas is likely in the case of supermassive {BH} (SMBH) LISA sources, since the galaxy mergers responsible for the binary itself can trigger large inflows of gas towards the central regions{, because of both tidal torques (due to gravitational forces; see, e.g. \citealt{Barnes_Hernquist_1996,Hopkins_Quataert_2010}\hypersetup{citecolor=blue}; \citealt{2015capelo}) and hydrodynamical torques (e.g. large-scale ram-pressure shocks; \citealt{Barnes_2002,Capelo_Dotti_2017,Blumenthal_Barnes_2018}\hypersetup{citecolor=blue})}. Furthermore, gas is thought to be one of the key agents that can aid binary hardening below pc scales \citep[see, e.g.][]{2020souzalima}. For sources embedded in a circumbinary disc, gas torques are well described as resulting from \textit{viscous} forces \citep[see, e.g.][]{2021doraziodisc}:

\begin{align}
    \label{eq:fdotvisc}
    \dot{f}_{\rm{visc}} =  A \frac{1+q}{q} \frac{\dot{M}}{M}f_{\rm r},
\end{align}

\noindent where $\dot{M}$ is the gas accretion rate onto the binary and $A$ is a dimensionless pre-factor that depends on disc properties. Following \citet{2021doraziodisc}, we adopt a value of $A = 3$, which is appropriate for binaries with $10^{-2} < q <1$. We scale the mass accretion rate with the \citet{1916eddington} limit:

\begin{align}
  \label{eq:mdot}
  \dot{M}= f_{\rm{Edd}} \dot{M}_{\rm{Edd}} =f_{\rm{Edd}} 4 \pi \frac{G M m_{\rm P}}{c\, \sigma_{\rm T}} \frac{f_{\rm{Edd}}}{\eta},
\end{align}

\noindent where $f_{\rm{Edd}}$ is the Eddington ratio, $m_{\rm P}$ the proton mass, $\sigma_{\rm T}$ the Thomson cross section, and $\eta = 0.1$ the radiative efficiency \citep[see, e.g.][]{2004marconi}. Typical values for $f_{\rm{Edd}}$ in active galactic nuclei at $z\sim 2$ range from $ 10^{-3}$ to $ 1$, peaked around the commonly assumed value of $\sim 10^{-1}$ \citep[see, e.g.][]{2015suh}. Larger Eddington ratios are thought to be common at higher redshifts \citep[][]{2010willott}.\newline

The {gravitational influence} of a third body can induce \textbf{tidal fields} and/or \textbf{linear accelerations} that affect the orbit of the inner binary {and can produce subtle modifications to the source's GW emission \citep[see, e.g.][]{2017Bonetti,2019alejandro,2019randall,2021alejandro,2022liu,2022xuan}\hypersetup{citecolor=blue}}. {As in the rest of this work, we will limit our analysis to the lowest-order effect on the inner binary's frequency evolution and also assume that the latter is the sole source of GWs. In a hierarchical triplet, the average effect of the third body is to induce} a tidal term $E_{\rm t}$ which modifies the {inner binary's} binding energy $E_{\rm b}$ \citep[see, e.g.][for a {recent} PN treatment of the three-body problem]{2021will}. Dimensionally, the ratio between the energies reads

\begin{align}
    \frac{E_{\rm{t}}}{E_{\rm b}} \sim B \frac{m_3 a^3}{M r^3},
\end{align}

\noindent where $a$ is the inner binary's separation, $r \gg a$ is the distance to the third body, $m_3$ its mass, and $B$ is a dimensionless pre-factor of order unity that depends on the configuration of the system, which we set equal to one. As shown in \cite{2021ii2}, the tidal term affects the inspiral rate of the inner binary by a factor proportional to $E_{\rm t}/E_{\rm b}$, modifying its frequency evolution:

\begin{align}
    \label{eq:fdottide}
    \dot{f}_{\rm{tot}}= \dot{f}\left(1 - 4 \frac{E_{\rm{t}}(f_{\rm r})}{E_{\rm{b}}(f_{\rm{r}})} \right).
\end{align}

\begin{figure}
    \centering
    \includegraphics[scale=0.65]{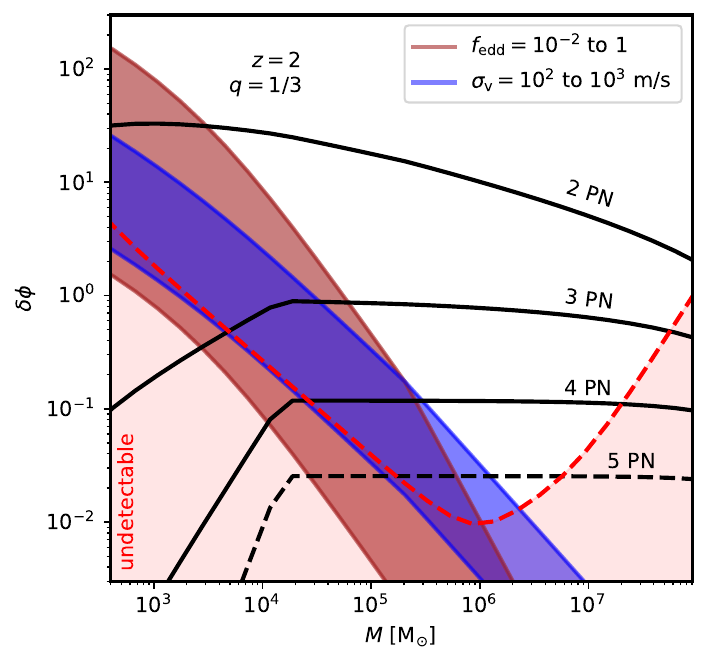}
    \vspace{-10pt}
    \caption{Phase shift caused by adding higher-order {phenomenological} PN terms (black lines, Eq.~\ref{eq:fdotphenom}) or environmental effects (brown and blue areas, Section~\ref{sec:env}) to the frequency evolution of sources of different mass. The red line denotes the minimum detectable phase shift considering the SNR of the source, as estimated by Eq.~\eqref{eq:mindeph}.}
    \label{fig:deph}
\end{figure}

A third body {of mass $m_3$ at a distance $r$ also induces an acceleration of the inner binary's centre of mass}, which can produce a time-dependent change in the peculiar velocity of the {GW} source. {While a constant Doppler shift is degenerate with redshift, a peculiar acceleration along the line of sight can cause a time variation in the Doppler shift which, if integrated over an observation time, causes a non-degenerate shift in the source's GW phase}. {We model this effect by Doppler shifting the frequency by an amount proportional to the time-dependent line-of-sight velocity $V_{\rm p}(t)$ which is solely caused by the third body}:

\begin{align}
    f \to f\left(1 +  \frac{V_{\rm p}({t})}{c}\right),
\end{align}

\noindent {where $V_{\rm p}(t) = (G m_3/r^2) t$. We can then integrate the evolution equations as usual to obtain an accumulated, non-degenerate dephasing caused by the centre of mass acceleration.} We {parametrize} this effect by considering the line-of-sight velocity, $\sigma_{\rm v}$, reached after a time $T$ of acceleration:

\begin{align}
    \label{eq:sigv}
    \sigma_{\rm v} = \frac{G m_3}{r^2}T.
\end{align}

In our calculations, $\sigma_{\rm{v}}$ therefore represents the maximum value of $V_{\rm p}(t)$ reached within an observation window. Typical values for $\sigma_{\rm v}$ are

\begin{align}
\frac{G m_3}{r^2} T \approx 4.4 \times \left(\frac{m_3}{10 \, \rm{M}_{\sun}} \right) \left( \frac{{10^{-4}} \, \rm{pc}}{r} \right)^2\left(\frac{T}{\rm{yr}} \right)\, \left[\frac{\rm{km}}{\rm{s}} \right].
\end{align}

\begin{figure}
    \vspace{-2.5pt}
    \centering
    \includegraphics[scale=0.65]{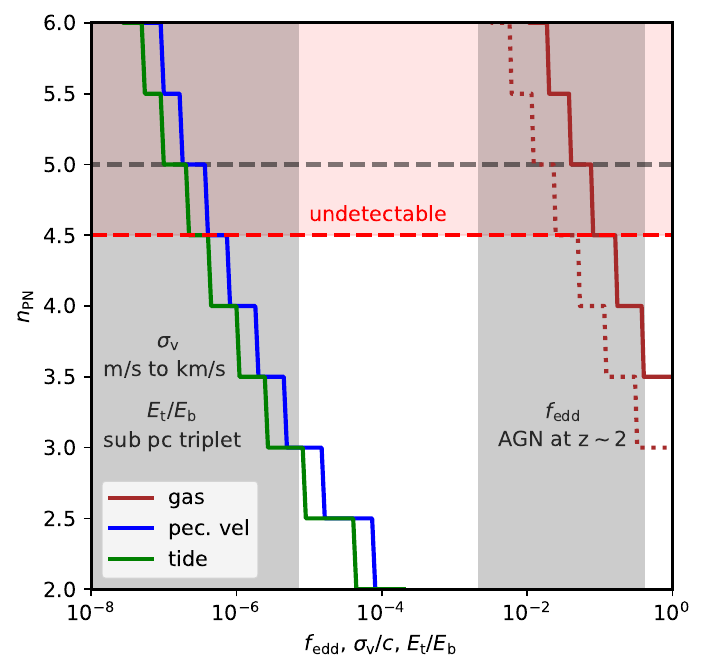}
    \vspace{-10pt}
    \caption{{Comparative} PN order of several environmental effects on a LISA source at $z=2$, with {total} mass $10^5$ M$_{\sun}$  and mass ratio 1/3 (solid) or 1/10 (dotted), as a function of their respective dimensionless variables. The curves are determined by comparing the amount of dephasing caused by environmental effects versus several high PN order contributions to Eq.~\eqref{eq:fdotphenom}. The gray areas highlight regions that are typical in a galactic centre. The red dashed line denotes the smallest phase shift detectable (in terms of its effective PN order) considering the SNR of the source, while the gray dashed line indicates the PN order of currently available waveforms.}
    \label{fig:nPN}
\end{figure}

Both unmodelled velocities of the order of km~s$^{-1}$ and tidal deformations of the order $E_{\rm t}/E_{\rm b}\sim 10^{-6}$ can be produced in an astrophysical context: the presence of a heavy star or stellar-mass BH in the innermost $\sim 10^{-2}$ to $\sim 10^{-4}$ pc of a nuclear cluster is a likely consequence of relaxation and mass segregation \citep[see, e.g.][]{1977bahcall,2009hopman,2022itai} or the presence of a large-scale accretion disc \citep[see, e.g.][]{2004goodman,2007levin,2016bellovary}. Similar values can be produced by pc-scale SMBH triplets, another channel that can produce hard SMBH binaries despite possible angular momentum barriers or other delays \citep[][]{2019bonetti}.

\vspace{-10pt}
\subsection{Methodology}

{Having set up our models for both vacuum waveform templates and environmental effects, we devise a simple strategy to compare the importance of typical environmental perturbations to the accuracy of an arbitrarily precise vacuum template.} We compare contributions to the total phase {of the GW signal} produced by additional PN orders and environmental effects by means of Eq.~\eqref{eq:totalphase} and the various forms of Eq.~\eqref{eq:fdot} discussed in the previous section. Every GW source is integrated from the time it enters the LISA band for a 4{-}yr period or{, if it occurs first, until it reaches a separation of $12 G M/c^2$}. An example of our computations can be seen in Figure~\ref{fig:deph}, in which we plot PN contributions to the GW phase for a range of possible sources, as well as the detectability criterion defined by Eq.~\eqref{eq:mindeph}.

\begin{figure*}
    \centering
    \includegraphics[scale=0.55]{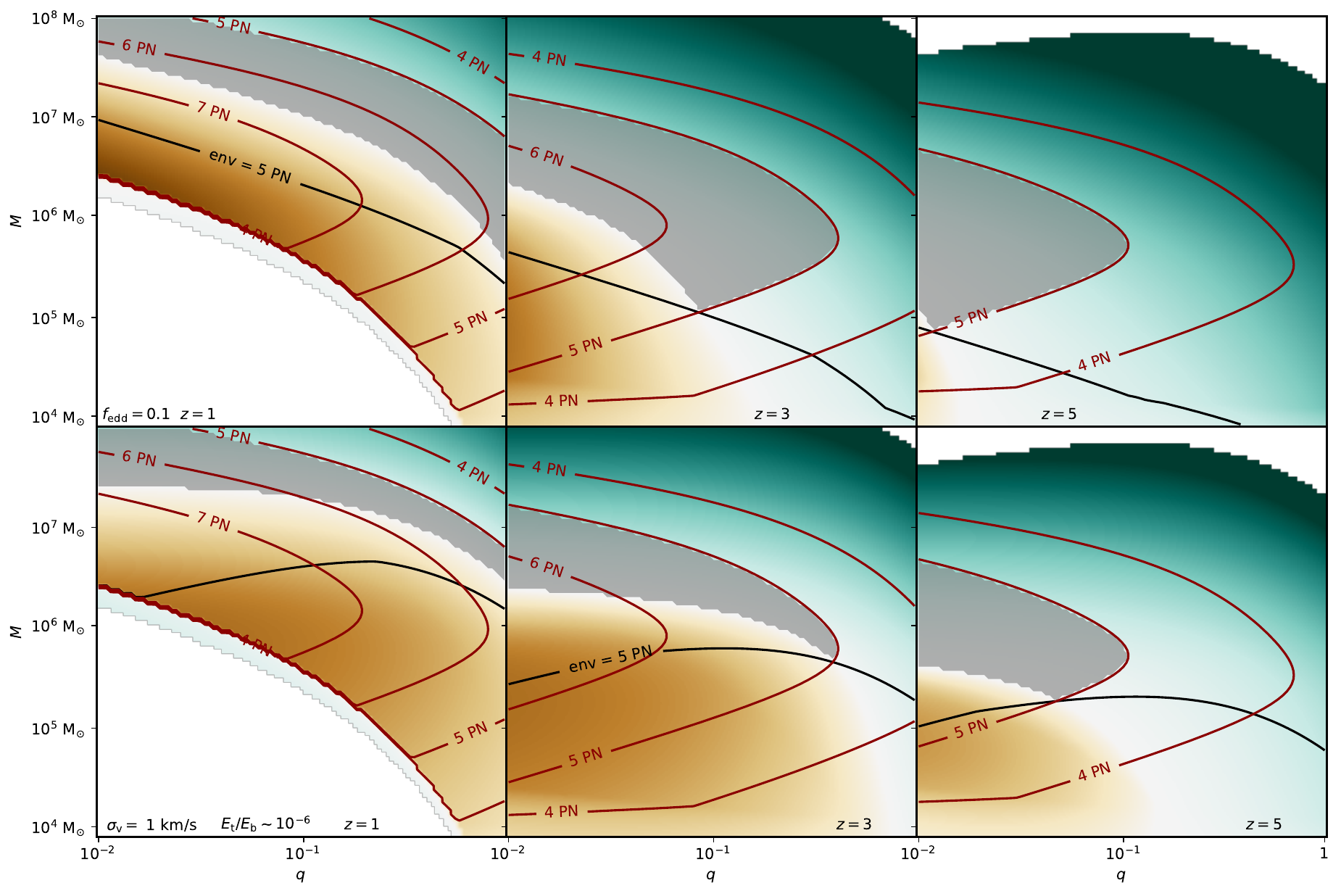}
    \vspace{-10pt}
    \caption{Parameter space (total mass and mass ratio) of LISA comparable-mass SMBH mergers at different redshifts -- $z = 1$ (right-hand panels), 3 (middle panels), and 5 (left-hand panels) -- including the effect of viscous torques (top panels) or a third body (bottom panels). Environmental perturbations are detectable (according to Eq.~\ref{eq:deccriterion} and imposing $\delta{\rm SNR} = 8$) in the brown regions, whereas the reverse is true in the blue regions. The darkness of the colour is a qualitative indication of the effective PN order of the environmental effect. The red contours denote the maximum detectable PN order contribution to the total phase of the source. The regions highlighted in gray satisfy the following criteria: 1) environmental effects are not detectable; and 2) contributions due to PN orders higher than five are detectable. The black line denotes the boundary at which the environmental influence is comparatively of 5th PN order.}
    \label{fig:phasespace}
\end{figure*}

We {term} an environmental perturbation to be of `{comparative} $n$-th PN order' if it produces a phase shift comparable to the $n$-th PN correction in Eq.~\eqref{eq:fdotphenom}. In Figure~\ref{fig:nPN}, we show the {comparative} PN order of selected environmental effects for {two} particular source{s} of GWs at $z=2$ with a {binary's total} mass of $10^5$ M$_{\sun}$ and a mass ratio $q = 1/3$ {and 1/10}, as a function of the dimensionless variables $f_{\rm{Edd}}$, $\sigma_{\rm v}/c$, and $E_{\rm t}/E_{\rm b}$. The {comparative} PN order of environmental effects strongly depends on the value of these parameters. For realistic values of $f_{\rm{Edd}} \sim 0.1$ and $\sigma_{\rm v}/c \sim E_{\rm t}/E_{\rm b} \sim 10^{-6}$, it tends to be between the 4-th and the 6-th PN order, comparable to the benchmark 5-PN precision of available waveform templates. {Note how very recent work on PN waveforms \citep[][]{owen23}\hypersetup{citecolor=blue} has determined that the truncation of 5PN terms will lead to a systematic bias in the parameter estimation of generic vacuum sources. By analogy, mis-modelling  environmental effects of comparative 5PN order is expected to lead to similar bias.}

\vspace{-10pt}
\section{Vacuum or Environment?}\label{sec:priorities}

\subsection{Comparable-mass mergers}

We apply our methodology on BH binaries with {total} masses between $10^3$ and $10^8$ M$_{\sun}$, mass ratios between $10^{-2}$ and 1, and redshift between 0 and 5. Sources with an SNR < 8 {according to Eq.~\eqref{eq:SNR}} are automatically discarded. The results of our analysis are visualised in Figure~\ref{fig:phasespace}, in which we assume representative values of $f_{\rm{Edd}} = 0.1$, $\sigma_{\rm v} = 1$~km~s$^{-1}$, and $E_{\rm t}/E_{\rm b} = 10^{-6}$. {With the aid of Eq.~\eqref{eq:deccriterion},} we show the detectability regions of environmental effects, represented by the brown (detectable) and blue (undetectable) areas. The red contours denote the maximum detectable PN order consistent with SNR limitations{, also according to Eq.~\eqref{eq:deccriterion}}. The regions highlighted in gray are defined by enforcing the following criteria: 1) environmental effects are not detectable; and 2) phase contributions due to PN orders higher than the 5 PN benchmark are detectable. The significant trends in the figure can be {roughly} summarised as follows:

\begin{itemize}

    \item {In {unequal-mass} binaries with total masses of order $\lsim 10^7/(1+z)^2$ M$_{\sun}$, environmental effects are likely to produce a larger phase contribution than the benchmark 5 PN precision.}

    \item {The accumulated phase of} heavy binaries ($\gsim 10^8/(1+z)$~M$_{\sun}$) can be adequately modelled with available waveforms due to their lower SNRs {and total accumulated phase}.
    
    \item Sources with the highest SNRs land around a total mass of $\sim 10^6$ M$_{\sun}$ and mass ratio of $\lsim 0.2$. In this range, environmental {phase contributions} are {relevant} at low redshift ($z \lsim 2$), while high-order vacuum {phase contributions} dominate at $z \gsim 4$.
    
\end{itemize}

Varying the choice of $f_{\rm{Edd}}, \sigma_{\rm v}$, or $E_{\rm t}/E_{\rm b}$ strongly affects the results, as suggested by Figure~\ref{fig:nPN}. For example, a larger Eddington fraction implies detectable {phase contributions} even at higher redshift \citep[see][for a detailed analysis]{garg2022}.

The results of Figure~\ref{fig:phasespace} must be interpreted in light of the expected merger rates of BH binaries. Several estimates suggest that lighter sources ($\lsim$ few $10^5$~M$_{\sun}$) at relatively low redshift ($\lsim 3$) are expected to dominate event rates, although significant variation can be caused by changing seeding prescriptions \citep[see, e.g.][]{2005rhook,2005sesana,2007sesana,2020volonteri}. If these estimates are accurate, they imply that a majority of LISA sources will likely fall in the brown regions of Figure~\ref{fig:phasespace}, strongly suggesting that the study of environmental effects should {also become} priority when it comes to LISA SMBH binaries.

\vspace{-10pt}
\subsection{{A comment on} extreme mass-ratio inspirals}

A second class of sources requiring complex waveforms are extreme mass-ratio inspirals (EMRIs). In the standard dynamical (dry) formation channel, stellar-mass BHs are scattered onto low peri-apsis orbits around SMBHs by relaxation processes \citep[see, e.g.][]{2003hp,2004pau,2013merrit,2022aceves} and are able to complete $\sim 10^5$ orbits only a few Schwarzschild radii above the event horizon of the primary BH. Because of their high initial eccentricity and rapid circularisation, EMRIs can ``skip'' over low-frequency GW emission: they enter the LISA band directly in a regime where environmental perturbations are completely negligible.

However, an alternative formation pathway can potentially produce similar expected EMRI rates. In the so-called active galactic nucleus (wet) channel, compact objects align or form within the central SMBH's accretion disc \citep[see, e.g.][]{1995rauch,2021wet,2022derdz}. They are dragged into the LISA band by means of gas torques, producing low-eccentricity EMRIs. They will likely enter the LISA band at lower frequencies, damping out high-order PN effects and reducing the overall SNR. While this might be a blow to precision general-relativity measurements, gas-embedded EMRIs could be powerful probes of accretion disc physics, enabling measurements inaccessible to electromagnetic instruments (see, e.g. \citealt{1993chakrabarti,1995ryan,2007levin,kocsis,2008barausse,2014barausse}\hypersetup{citecolor=blue}; \citealt{2019andrea,2021andrea,2022zwick,2022speri}; \citealt{2022destounis,2022polcar}\hypersetup{citecolor=blue}).

Because of the high uncertainties in the event rates, a choice between prioritising vacuum templates is premature. In any case, the detection of EMRIs is likely to provide a wealth of scientific observations, be it in fundamental physics \citep[see, e.g.][]{2017science,2022fplisa} or astrophysics.

\vspace{-10pt}
\section{Setting the priorities}\label{sec:conclusion}

In short, our analysis suggests that the majority of {massive binary} mHz {GW} sources are likely to populate the regions of parameter space in which 1) currently available waveforms can already {precisely match the phase of realistic signals if SNR limitations are considered}, or 2) {environmental phase contributions are likely dominant over vacuum corrections of order higher than the current benchmark value of 5 PN}.

The aim of developing high PN order templates is to perform the precision measurements required to test fundamental physics, in particular modifications to general relativity. Unless the environment is properly modelled, these measurements {will most likely} require special sources (denoted by the gray areas in Figure~\ref{fig:phasespace}) in which the SNR is sufficiently high and the environment is negligible. Additionally, most of the astrophysical information regarding the BH binary itself (mass, mass ratio, component spin, and orientation) can, in principle, be recovered with lower PN order phenomenological templates, since even complex dynamics such as spin-orbit and spin-spin couplings {produce phase shifts at much less than} $5$th PN order. On the other hand, further development on the environmental side would 1) reduce the possibility of unknown biases and 2) allow to constrain the astrophysical surrounding of GW sources. The latter is especially promising if richer environmental signatures are also taken into account. \newline \newline
Led by these considerations, we argue that \textit{systematically including environmental effects} in waveform templates should take priority with respect to further increasing the accuracy of {inspiral} vacuum templates. If the goal is to maximise the science yield of future missions, the community could be better served by shifting the focus from the source of GWs to its surroundings. 

\vspace{-15pt}
\section*{Acknowledgements}
The authors acknowledge support from the Swiss National Science Foundation under the Grant 200020\_192092.

\vspace{-15pt}
\section*{Data availability}
The data underlying this article will be shared on reasonable request to the authors.

\vspace{-15pt}
\scalefont{0.94}
\setlength{\bibhang}{1.6em}
\setlength\labelwidth{0.0em}
\bibliographystyle{mnras}
\bibliography{references}
\normalsize

\bsp 
\label{lastpage}
\end{document}